\documentclass[twocolumn,prl,aps,epsf,showpacs]{revtex4}

\usepackage{amssymb,epsf}
\usepackage{graphicx}
\usepackage{epstopdf}

\begin{document}

\author{C.R. Dean$^{1}$, B.A. Piot$^{1}$, G. Gervais$^{1}$, L.N. Pfeiffer$^{2}$ and K.W. West$^{2}$ }

\address{$^{1}$Department of Physics, McGill University, Montreal, H3A 2T8, CANADA}
\address{$^{2}$Bell Laboratories, Alcatel-Lucent Inc., Murray Hill, NJ 07974 USA}



\title{Current-induced nuclear-spin activation in a two-dimensional electron gas }

\begin{abstract}

Electrically detected nuclear magnetic resonance  was studied in detail in a two-dimensional electron gas
as a function of  current bias and temperature. We show that applying a  relatively modest 
dc-current bias, $I_{dc}\simeq 0.5$ $\mu A$, can induce a re-entrant and even enhanced nuclear spin signal compared with the 
signal obtained under similar thermal equilibrium conditions at zero current bias.
Our observations suggest that dynamic nuclear spin polarization by small current flow is possible 
in a two-dimensional electron gas, allowing for easy manipulation of the nuclear spin by simple
switching of a dc current.

\end{abstract}
 \maketitle

One of the main problems plaguing the implementation of quantum information processing schemes in GaAs/AlGaAs quantum dots is the rather short coherence lifetimes of the quantum states; typically lasting only $T^{*}\sim ns$  \cite{Petta05}.  An important source of decoherence in these systems is the electron-nuclear spin interaction resulting from the strong hyperfine coupling between the two-dimensional electron gas (2DEG) and the surrounding nuclei of the GaAs/AlGaAs substrate. While presenting a nuisance for electronic-based quantum information devices, the nuclear spins in semiconductors have themselves been proposed as an alternate candidate for  quantum information carriers\cite{Salis01,Loss02}, and their coherent manipulation by pulsed techniques has been demonstrated experimentally \cite{Yusa05}.  This is particularly appealing due to their high degree of isolation and thus their resilience to decoherence, albeit  the polarization of small ensembles of nuclear spins, as well as their addressing and initializing remain formidably challenging experimentally. 



It has been well demonstrated that the electron-nuclear hyperfine coupling in GaAs-based systems can be exploited for the electrical detection of nuclear magnetic resonances (NMR) in the quantum Hall regime \cite{Dobers:PRL:1988, Kronmuller:PRL:1999,Desrat:PRL:2002,Gervais:PRL:2005,Hirayama}. For use in quantum information however,  
the challenge is to both enhance the polarization so as to be detectable, and also to have fine control over the full degree of polarization.  The nuclear spin in GaAs/AlGaAs can in principle be electrically resolved whenever the average nuclear polarization in a magnetic field is sufficiently large so that electronic transport  becomes sensitive to a small change in nuclear spin orientation through the electronic Zeeman energy. To achieve increased sensitivity, the nuclear polarization can be {\it dynamically enhanced} through electron-nuclei flip-flopping processes with, for example, electron spin resonance \cite{Dobers:PRL:1988},  edge states in point contact devices \cite{Wald94}, and recently with  single-electron transport in quantum dots \cite{Ono04,Petta08}. Since the nuclear polarization in the presence of a magnetic field follows a Boltzmann temperature distribution, becoming substantial at temperatures below $T\lesssim 100$ mK, the nuclear polarization can instead be enhanced simply by reducing the sample temperature, without recourse to any dynamic polarization schemes \cite{Kronmuller:PRL:1999,Desrat:PRL:2002,Gervais:PRL:2005}.  While conceptually simple, thermally inducing variations of the nuclear spin lacks the control required to tune the the nuclear polarization within time-scales on the order of the spin-relaxation time, $T_{1}$. In this Letter, we perform a two-dimensional frequency-current mapping of the electrically detected NMR signal in the low temperature regime, and find strong evidence that even a modest electrical dc-current can activate the nuclear spins in a bulk 2DEG.  This opens new avenues towards the complete all-electrical control and measurement of the nuclear spins in GaAs/AlGaAs semiconductors.

In the quantum Hall regime of a 2DEG, the Zeeman energy gap in the non-interacting electron picture is given by $\Delta_{Z}=g^{*}\mu_{B}(B+B_{N})$, 
where $g^{*}$ is the effective electron g-factor, $\mu_{B}$ is the Bohr magneton, $B$ is the applied dc magnetic field, and $B_{N}={\cal A}\left<I_{z}\right>/g^{*}\mu_{B}$  is the Overhauser field  which depends on the strength of the hyperfine coupling constant ${\cal A}$ and
the nuclear spin polarization $\left<I_{z}\right >$.   Applying transverse RF radiation at the nuclear resonant frequency destroys the nuclear polarization which in turn decreases $B_{N}$, thereby altering the electronic Zeeman gap.  Due to the negative g-factor in GaAs ($g^{*}$=-0.44), $B_{N}$ opposes $B$ so that destroying the nuclear polarization increases the overall Zeeman energy. This modification of the Zeeman gap by RF-radiation can be electrically detected since, in the thermally activated quantum Hall regime, the longitudinal resistance is sensitive to the gap energy, $R_{xx}\propto e^{-\Delta/2k_{B}T}$.  This holds true for  most filling factors where a minimum in $R_{xx}$ is observed at resonance, exhibiting a Lorentzian lineshape provided the RF-frequency is swept (in a continuous-wave experiment)  at a rate slower than the spin-lattice relaxation rate, $\tau ^{-1}\ll 1/T_{1}$. The magnitude and position of the NMR response yields a wealth of information about the spin state of both the nuclei and electron gas.

\begin{figure}[t]
	\begin{center}
		\includegraphics[width=1\linewidth,angle=0,clip]{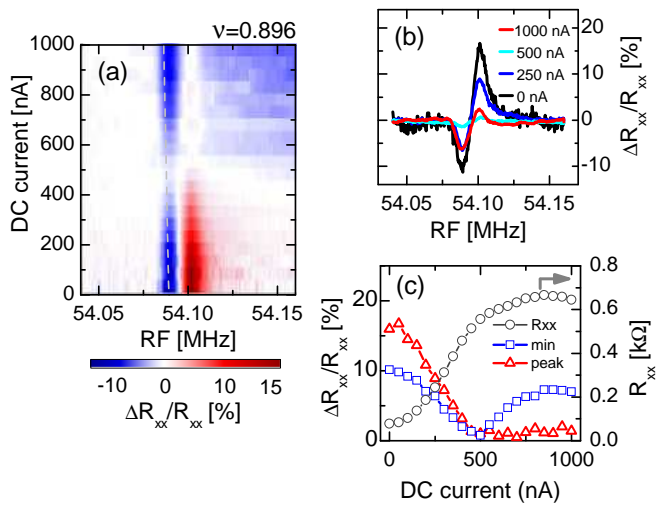}
		\caption{ (a) 2D contour map of the NMR lineshape versus dc current bias at $\nu=0.896$ 
and $T_{e}=34mK$ at $I_{dc}=0$ nA. Dashed line indicates the estimated magnet drift during the scan. (b) Selected traces from the contour plot in (a). 
(c) Normalized minima  (open squares) and peak (open triangles) versus dc current bias, plotted 
along with the corresponding off-resonant background resistance (open circles).  }
		\label{fig2}
	\end{center}
\end{figure}

This simple picture of electrically detected NMR in the quantum Hall regime breaks down {\it de facto} in the first Landau level where in the vicinity of  the  $\nu=1$ quantum Hall state, an ``anomalous dispersive'' lineshape is observed \cite{Desrat:PRL:2002}.  At the on-resonance condition $f_{RF}=f_{Larmor}$,  the usual resistance minimum is followed by a secondary resistance peak of unknown origin  \cite{Desrat:PRL:2002,Tracy:PRB:2006, Kodera:PhysStatSol:2006, Kawamura:PhysE:2008}. A similar dispersive signal has recently been identified at other filling factors in the lowest Landau level \cite{Stern:PRB:2004, Gervais:PRB:2005}, and a peak-only response has been reported at $\nu$=1/2 \cite{Stern:PRB:2004, Tracy:PRL:2007}.  Fig. \ref{fig2}a shows a two-dimensional frequency-current contour plot of the electrically detected NMR signal measured in our sample at $\nu$=0.896, with some individual spectra shown in Fig. \ref{fig2}b.  At zero dc current bias, where the electrically detected NMR technique has previously been studied, we observe the typical dispersive lineshape reported elsewhere.

\begin{figure}[t]
	\begin{center}
		\includegraphics[width=0.9\linewidth,angle=0,clip]{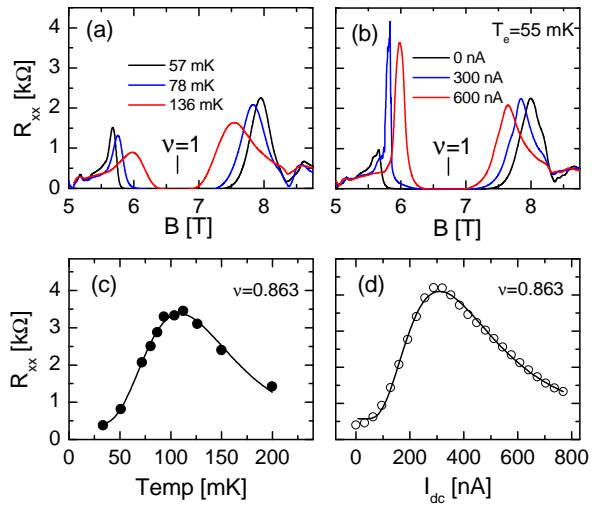}
		\caption{(a,b) Variation in the longitudinal resistance, $R_{xx}$, around the $\nu$=1 quantum Hall state due to (a) sample heating and (b) applying a dc-current bias.  (c,d) $R_{xx}$ versus sample temperature and applied dc-current, respectively, at filling $\nu$=0.863 ($B$=7.800~Tesla).  Data in (c) and (d) were acquired at fixed field, on a separate cooldown from the data in (a) and (b).}
		\label{fig1}
	\end{center}
\end{figure}

Our sample is a 40~nm wide GaAs/AlGaAs quantum well containing a 2DEG with a measured mobility $\mu\sim16.6\times10^{6}~\text{cm}^{2}/\text{Vs}$ and electron density $n_{e}\sim1.6\times10^{11}$~cm$^{-2}$. NMR measurements were performed at fixed field by continuously shining transverse RF radiation at constant power (CW mode) while sweeping the RF frequency through the Larmor resonance condition.  Resonance was observed via electrical transport measurements using a standard lock-in technique at low frequency ($\sim10$~Hz) and small excitation currents ($\sim 10$~nA). All NMR  measurements are reported for the $^{75}As$ nuclei only; measurements of the Ga isotopes are expected to give the same result \cite{Desrat:PRL:2002, Kodera:PhysStatSol:2006}.  The temperatures quoted are the electron temperature, calibrated against a CMN thermometer and superconducting fixed points, and corrected for non-resonant RF heating.

The most striking feature of the data shown in Fig. 1a is the variation in the NMR signal with increasing dc current bias.  
The data was acquired by stepping $I_{dc}$ in regular-intervals and sweeping the RF frequency through resonance at each current value.  During each scan,  the magnet was held persistent.  The frequency sweep time at each interval, plus a $\sim$10 minutes pause between intervals, gave a total scan time of $\sim$10~hrs for the whole plot.  The dashed line indicates the expected magnet drift over this scan time, estimated by repeatedly measuring the NMR signal over a similar time-scale. From the contour plot, we observe the following evolution of the NMR signal:  {\it i)} both the minimum and peak features initially diminish, nearly vanishing at $I_{dc}$$\sim$500~nA;  {\it  ii)} as $I_{dc}$ is further increased the minimum re-appears and gains in intensity with $I_{dc}$.   Our basic result, {\it i.e.} the re-entrance and the strengthening of the signal above a critical dc-current value  $I_{dc}\gtrsim I_{c}$ has been reproduced at other filling fractions in the flank of the $\nu$=1 quantum Hall state.

Application of a dc current causes heating of the electron gas.  The initial diminishing of the NMR signal therefore could be understood as a consequence of associated thermal destruction of the nuclear polarization.
The thermal distribution of the nuclear spin magnetization  should roughly follow that of a Curie law $\left <I_{z}\right >\propto B/T$ when $\mu_{N} B << k_{B}T$,  {\it i.e} in the high temperature/low field limit.  Increasing the sample temperature should therefore reduce the nuclear spin polarization and restore the electronic Zeeman gap to its `nuclear spin-free'  value. Consequently, the electrical NMR signal strength $\Delta R_{xx}/R_{xx}$  is expected to vanish monotonically to zero with increasing temperature, consistent with the initial trend in Fig. 1a.  However, in this view, the subsequent reentrance of the NMR signal at higher dc current values is unexpected. The re-emergence of the NMR signal at increased dc current bias, where thermal effects are expected to further destroy the nuclear polarization, suggests enhancement of the nuclear spin state by dynamic nuclear polarization (DNP) \cite{Kawamura:ApplPhysLett:2007,Kawamura:PhysE:2008,Zhang:PRL:2007} may be at play.

\begin{figure}[t]
	\begin{center}
		\includegraphics[angle=0,clip]{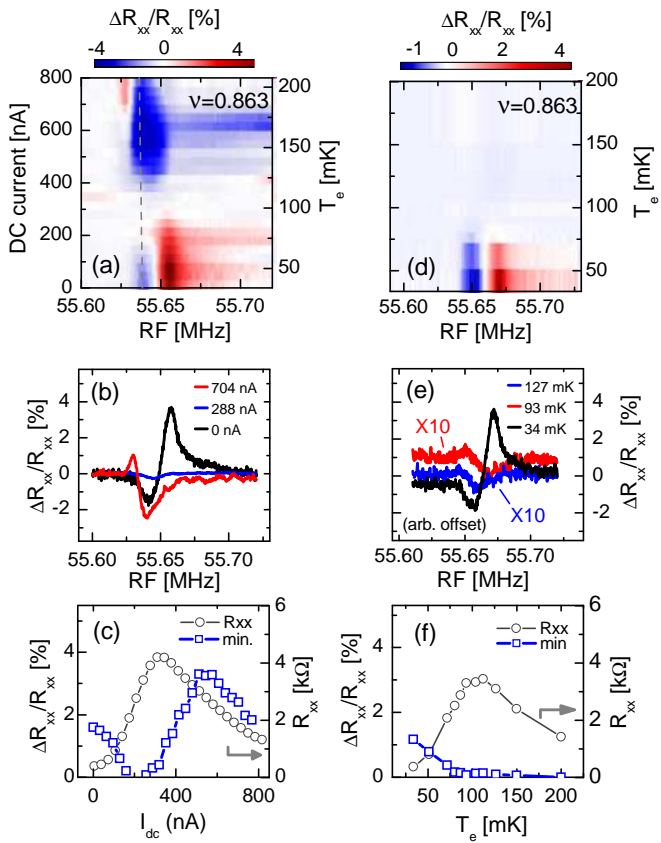}
		\caption{(a) Contour plot of the dispersive lineshape versus dc-current bias at $\nu$=0.863. The right axis shows the corresponding electron temperature. (b) Selected traces from the data in (a). (c) Normalized NMR minimum (open squares)  together with the off-resonant background resistance (open circles).  (d)-(e) Same plots as in (a)-(b) but resulting from varying the mixing chamber temperature with no dc-current applied to the sample. Dashed line in (a) indicates magnet drift during scan.}
		\label{fig3}
	\end{center}
\end{figure}

To deconvolve contributions from the dc current and thermal effects we compare the result of sample heating due to application of a dc current, versus heating the sample directly by raising the fridge temperature. Fig. \ref{fig1}a shows the effect of increasing the electronic temperature on the $\nu=1$ quantum Hall state  when the refrigerator temperature was increased.  This causes a convergence of the resistance peaks on either side of the minimum.  As a result, at fixed filling factor (Fig. \ref{fig1}c), $R_{xx}$ first increases with temperature ($dR_{xx}/dT>0$), only to decrease with further heating ($dR_{xx}/dT<0$).   Applying a dc-current bias $I_{dc}$ through the electrical leads contacted to the 2DEG gives a nearly identical trend, as shown in Fig. \ref{fig1}b,d. By comparing the magnetoresistance variation resulting from heating the sample via the fridge with that of  applying the dc-current, the electron heating due to $I_{dc}$ can be estimated \cite{Wei:PRL:1988, Chow:PRL:1996}.  The correspondence between $R_{xx}(I_{dc})$ and $R_{xx}(T)$ in the $\nu\sim 0.86$ region in Fig. \ref{fig1} indicates that likewise $dR_{xx}/dI_{dc}\propto dR_{xx}/dT$. This suggests that the re-entrance of the NMR signal we observe in Fig. 1 is not simply due to a crossover in the sign of $dR_{xx}/dT$ \cite{Tracy:PRB:2006} since the signal is observed to vanish at $I_{dc}\simeq 500$ nA and then reappear where $R_{xx}$ is still increasing with $I_{dc}$, {\it i.e.} in a region where $dR_{xx}/dT\propto dR_{xx}/dI_{dc}$ is always positive.

In Fig. \ref{fig3} we show a direct comparison of the NMR signal evolution obtained by varying  $I_{dc}$ (Fig. \ref{fig3}a-c) versus varying the fridge temperature with no dc current applied (Fig. \ref{fig3}d-f) at a fixed filling factor $\nu=0.863$. This data was acquired at the same filling factor as the data in Fig. \ref{fig1}, allowing us to calibrate and compare the electron heating due to the dc current.  When heating the sample by increasing the fridge temperature,  the NMR signal initially diminishes with increasing $T_{e}$,  and approaches a vanishingly small signal intensity at electron temperature $T_{e}$$\sim$100~mK. This behaviour is similar to that observed at low current bias prior the re-rentrance of the NMR signal, indicating that in both cases, the weakening of the NMR signal in this regime is  likely a purely thermal effect, {\it i.e.} due to a decrease in nuclear spin polarization as the temperature is increased.  However, unlike when the dc-current bias is applied, the signal does not show any re-entrance upon further heating of the substrate. Closer examination of the high-temperature curves, plotted in Fig. \ref{fig3}e, reveals that the signal does persist to high temperature but remains very small, barely perceptible above the background noise. To emphasize this, the 93~mK and 127~mK traces in Fig. \ref{fig3}e were multiplied by a factor of ten and offset vertically to allow for a direct comparison with the initial signal taken near base-temperature. The NMR signal produced by a dc current bias above $I_{c}$ exceeds that of the thermally induced (at same electronic temperature) by a factor of $\sim 50$. 


At $I_{dc}$$\sim$650~nA, where we estimate the electron temperature to be $\sim$170~mK,  the magnitude of the normalized NMR minimum is nearly twice its initial value (open squares in Fig. \ref{fig3}c).  By contrast, when heating the sample to this temperature by varying the substrate temperature, there is effectively no measurable NMR response (open squares in Fig. \ref{fig3}f). The magnitude of the NMR signal is in principle limited by the nuclear polarization if thermally populated.  A persistence of the signal to large dc-current values could be explained by the current-induced heating predominantly affecting the electron temperature without changing much the nuclear temperature.  This might be expected since at very  low temperatures, as in our experiment,  the electrons may not be that well thermally coupled to the lattice. However, the observed \textit{enhancement} of the NMR minimum, increasing by nearly a factor of two  for $I>I_{c}$  is strong evidence that the applied dc current actively {\it enhances} the nuclear polarization even beyond its thermal equilibrium value at  $T\simeq 20$ mK. 

We also note in Fig. \ref{fig3}a that the peak signal re-emerges at high dc current bias but appears downshifted, occurring at a lower frequency position than the  minimum, consistent with the NMR lineshape inversion reported previously \cite{Tracy:PRB:2006, Dean:PhysicaE:2008}. The origin of the peak in the dispersive lineshape remains unknown, making it difficult to understand the mechanism that causes an apparent shift in the peak position with applied dc current bias. However, this observation rules out the possibility that the enhanced minimum at high $I_{dc}$ is somehow the result of the initial minimum and peak signals collapsing onto a single response, and therefore further supports our interpretation that the signal enhancement results from a current induced DNP process.

The re-entrance, persistence and even the strengthening of the NMR signal with applied dc current
constitute a collective set of evidence for  dynamic nuclear polarization (DNP) induced by the current. This is 
consistent with recent reports of a current-induced DNP near $\nu=1$ \cite{Kawamura:ApplPhysLett:2007, Kawamura:PhysE:2008, Zhang:PRL:2007}. In DNP, nuclear spin polarization is induced through non-thermal dynamical processes where electron-nucleus 
flip-flop via  the hyperfine hamiltonian ${\cal H}={\cal A}\vec{I}\cdot\vec{S}=\frac{\cal{A}}{2} [I_{+} S_{-}+I_{-}S_{+}]+{\cal A}I_{z}S_{z}$.
For this to occur, spin-flip scattering of electrons is necessary, which is known to be occurring between spin-resolved quantum Hall edge channels, or with domain structure of different spin configurations. Recent work by Kawamura {\it et al.}\cite{Kawamura:ApplPhysLett:2007} suggested that  DNP is possible in principle in bulk 2DEG provided the dc current
exceeded a threshold value $I_{c}\gtrsim 0.3-1.0 \mu A$. Their findings are fully consistent with ours, and support the view that DNP can occur with relatively small electrical currents.

In summary, we report on a dc-current induced re-entrance of the anomalous dispersive NMR lineshape observed in the vicinity of $\nu$=1 quantum Hall state. Comparing the effects of a dc-current bias with those of direct sample heating, we conclude that
two distinct mechanism are at play in the detection of the NMR signal electrically, one that is purely thermally activated and the other due to the current flow intensity.  Our observation of an increase of the NMR signal at large dc-current bias is strong evidence for a current-induced  nuclear spin enhancement by means of dynamic nuclear polarization. This effect is of high relevance to the control and  manipulation of small ensembles of nuclear  spins in semiconductors for applications such as information storage and encoding into a `nuclear spin memory'.

This work has been supported by the Natural Sciences and Engineering Research Council of Canada (NSERC), the Canada Fund for Innovation (CFI), the Canadian Institute for Advanced Research (CIFAR), FQRNT (Qu\'{e}bec) and the A. P. Sloan Foundation (G.G.). We thank  R. Talbot, R. Gagnon and J. Smeros for technical assistance,  as well as W. Coish for illuminating discussions. 

\end{document}